\documentclass[12pt]{article}

\usepackage{amssymb,amsmath,mathrsfs,epstopdf,slashed,color}
\usepackage{ifpdf}
\ifpdf
\usepackage{graphicx}
\usepackage{hyperref}
\usepackage{cite}
\else
\usepackage[dvipdfmx]{graphicx}
\usepackage[dvipdfmx]{hyperref}
\usepackage[numbers]{natbib}
\usepackage{hypernat}
\fi

\allowdisplaybreaks[1]

\makeatletter

\@addtoreset{equation}{section}
\makeatother

\newcommand{\Vol}{\mathcal{V}}

\DeclareMathOperator*{\re}{Re \,}
\DeclareMathOperator*{\im}{Im \,}

\title{\bf De Sitter Vacua from a D-term Generated Racetrack Uplift}
\author{Yoske Sumitomo, Markus Rummel}

\begin{document}

\begin{titlepage}

\setcounter{page}{0}
  
\begin{flushright}
 \small
 \normalsize
 KEK-TH-1755
\end{flushright}

\vskip 3cm
\begin{center}

{\Large \textbf{De Sitter Vacua from\\ a D-term Generated Racetrack Uplift}}

\vskip 2cm
  
{\large Markus Rummel${}^{1}$ and Yoske Sumitomo${}^{2}$, }
 
 \vskip 0.6cm

${^1}$Rudolph Peierls Centre for Theoretical Physics, University of Oxford,\\ 1 Keble Road, Oxford, OX1 3NP, United Kingdom\\
${}^2$ High Energy Accelerator Research Organization, KEK,\\ 1-1 Oho, Tsukuba, Ibaraki 305-0801 Japan\\

 \vskip 0.4cm

Email: \href{mailto:markus.rummel@physics.ox.ac.uk, sumitomo@post.kek.jp}{markus.rummel at physics.ox.ac.uk, sumitomo at post.kek.jp}

\vskip 1.0cm
  
\abstract{\normalsize
We propose an uplift mechanism using a structure of multi-K\"ahler moduli dependence in the F-term potential of type IIB string theory compactifications. This mechanism requires a D-term condition that fixes one modulus to be proportional to another modulus, resulting in a trivial D-term potential. De Sitter minima are realized along with an enhancement of the volume in the Large Volume Scenario  and no additional suppression of the uplift term such as warping is required. We further show the possibility to realize the uplift mechanism in the presence of more K\"ahler moduli such that we expect the uplift mechanism to work in many other compactifications.
}
  
\vspace{1cm}
\begin{flushleft}
 \today
\end{flushleft}
 
\end{center}
\end{titlepage}

\setcounter{page}{1}
\setcounter{footnote}{0}

\tableofcontents

\parskip=5pt

\section{Introduction}

Dark Energy is the dominant source causing the current accelerated expansion of the universe, as has been confirmed by observations~\cite{Riess:1998cb,Schmidt:1998ys,Bennett:2012zja,Ade:2013zuv}. Although there exist some possibilities explaining Dark Energy, a tiny positive cosmological constant would be the prime candidate, in perfect agreement with recent observations~\cite{Bennett:2012zja,Ade:2013zuv}.

If one wants to understand the purely theoretically origin of this cosmological constant, we should promote Einstein gravity to be consistent with its quantum formulation. String theory is quite motivated for this purpose as it is expected to provide the quantum nature of gravity as well as particle physics. A cosmological constant could be realized in the context of flux compactifications~\cite{Giddings:2001yu,Dasgupta:1999ss} of 10D string theories, where a vacuum expectation value of the moduli potential at minima contributes to the vacuum energy in a four-dimensional space-time universe. Since there exist many possible choices of quantized fluxes and also a number of types of compactifications, the resultant moduli potential including a variety of minima forms the string theory landscape (see reviews~\cite{Douglas:2006es,Grana:2005jc,Blumenhagen:2006ci,Silverstein:2013wua,Quevedo:2014xia,Baumann:2014nda}). 

Although there exist many vacua in the string theory landscape, when we naively stabilize the moduli and obtain the minima, negative cosmological constants seem likely to come by. Hence an 'uplift' mechanism from the negative vacuum energy keeping stability should be important to realize an accelerated expanding universe. Some possible ways of the uplift mechanism have been proposed in string compactifications.

\begin{itemize}

\item Explicit SUSY breaking achieved by brane anti-brane pairs contributes positively in the potential, and thus can be used for the uplift~\cite{Kachru:2002gs,Kachru:2003aw,Kachru:2003sx}. When the D3 brane anti-brane pairs are localized at the tip of a warped throat, the potential energy may be controllable due to a warping factor. As the uplift term contributes to the potential at ${\cal O}({\cal V}^{-4/3})$, which appears larger than the F-term potential for stabilization which is in general $\mathcal{O}(\ll \Vol^{-2})$, de Sitter ($dS$) vacua with tiny positive cosmological constant may be achieved as a result of tuning of warping. A caveat of this proposal is that the SUSY breaking term needs to compensate the entire Anti-de Sitter ($AdS$) energy, so it is an open question if the SUSY breaking term, originally treated in a probe approximation, can be included as a backreaction in supergravity appropriately.

\item As an alternative uplift mechanism, one may use the complex structure sector~\cite{Saltman:2004sn}. In the type IIB setup, the complex structure moduli as well as the dilaton are often stabilized at a supersymmetric point. Owing to the no-scale structure, the potential for the complex structure sector is positive definite $V_{\rm c.s.} = e^{K} |DW|_{\rm c.s.}^2 \sim {\cal O} (\Vol^{-2})$. So when we stay at the SUSY loci, the potential is given convex downward in general and thus tractable. However, if one stabilizes the complex structure sector at non-supersymmetric points, then there appears a chance to have a positive contribution in the potential without tachyons, that may be applied for the uplift with a tuning. See also recent applications of this mechanism~\cite{Danielsson:2013rza,Blaback:2013qza,Kallosh:2014oja}.

\item When we include the leading order $\alpha'$-correction coming of ${\cal O}(\alpha'^3)$ in the K\"ahler potential~\cite{Becker:2002nn} which breaks the no-scale structure, this generates a positive contribution in the effective potential if the Euler number $\chi$ of the Calabi-Yau is given by a negative value~\cite{Balasubramanian:2004uy}. This positive term can balance the non-perturbative terms in the superpotential such that stable $dS$ vacua can be achieved in this K\"ahler Uplift model~\cite{Westphal:2006tn,Rummel:2011cd,deAlwis:2011dp} (see also~\cite{Westphal:2005yz}). In the simplest version of the K\"ahler uplifting scenario, there is an upper bound on the overall volume of the Calabi-Yau such that one may worry about higher order $\alpha'$-corrections. However, this bound can be significantly relaxed when embedded in a racetrack model~\cite{Sumitomo:2013vla}.

\item It has been proposed that the negative curvature of the internal manifold may be used for $dS$ constructions as it contributes positive in the scalar potential~\cite{Silverstein:2007ac}. Motivated by this setup, there were many attempts constructing $dS$ vacua~\cite{Haque:2008jz,Flauger:2008ad,Danielsson:2009ff,Caviezel:2009tu,deCarlos:2009fq,deCarlos:2009qm,Dong:2010pm,Andriot:2010ju,Danielsson:2010bc,Danielsson:2011au,Danielsson:2012et,Danielsson:2012by,Blaback:2013ht}. Using the necessary constraint for the extrema~\cite{Hertzberg:2007wc,Wrase:2010ew} and for the stability~\cite{Shiu:2011zt}, we see that the existence of minima requires not only negative curvature, but also the presence of orientifold planes.

\item When the stabilization mechanism does not respect SUSY, D-terms can provide a positive contribution to the potential if the corresponding D7-brane is magnetically charged under an anomalous $U(1)$~\cite{Burgess:2003ic,Cremades:2007ig,Krippendorf:2009zza}. The potential of D-terms arises of order ${\cal O} ( {\cal V}^{-n})$ with $n\geq 2$, depending on the cycle that the D7-brane wraps. If we take into account the stabilization of matter fields having a non-trivial VEV, originating from fluxed D7-branes wrapping the large four-cycle, then the uplift contribution becomes ${\cal O} ({\cal V}^{-8/3})$~\cite{Cremades:2007ig}. So a relatively mild suppression, for instance by warping, is required for this volume dependence. See also recent applications to explicit scenarios in~\cite{Cicoli:2012vw,Cicoli:2013mpa} and also together with a string-loop correction in fibred Calabi-Yaus~\cite{Cicoli:2011yh}.

\item Recently, it has been proposed that a dilaton-dependent non-perturbative term can also work for the uplift mechanism toward $dS$ vacua~\cite{Cicoli:2012fh}. The non-perturbative term depends on both the dilaton and a vanishing blow-up mode which is stabilized by a D-term. Since the D-term turns out to be trivial at the minima due to a vanishing cycle, the non-trivial dilaton as well as the vanishing cycle dependence generate the uplift term within the F-term potential. In this setup, the given uplift term is proportional to $e^{-2b \langle s \rangle}/{\cal V}$. Although the volume does not suppress the uplift term so much, we may expect an exponential fine-tuning by the dilaton dependence to balance with the moduli stabilizing F-term potential.

\end{itemize}

In this paper, we introduce an uplifting term of the form $e^{-a_s \tau_s}/{\cal V}^2$, where $\tau_s$ is the volume of a small 4-cycle, which naturally balances with the stabilizing F-term potential in the Large Volume Scenario (LVS). The following ingredients are necessary for this mechanism:
\begin{itemize}
 \item one non-perturbative effect on a 4-cycle $D_2$ to realize the standard LVS moduli stabilization potential, 
 \item another non-perturbative effect on a different cycle $D_3$,
 \item a D-term constraint that enforces the volumes $\tau$ of the two 4-cycles to be proportional $\tau_s \sim \tau_2 \propto \tau_3$ via a vanishing D-term potential.
\end{itemize}
Hence, the minimal number of K\"ahler moduli for this uplifting scenario is $h^{1,1}_+=3$. At the level of the F-term potential the effective scalar potential reduces to the standard LVS moduli stabilization potential plus the mentioned uplifting term yielding metastable $dS$ vacua. The K\"ahler moduli are stabilized at large values avoiding dangerous string- and $\alpha'$-corrections. Compared to~\cite{Cicoli:2012fh}, the dilaton can take rather arbitrary values determined by fluxes as there is no tuning required to keep the uplifting term suppressed. Note that for $h^{1,1}_+=2$, a racetrack setup with two non-perturbative effects on one small cycle does not allow stable $dS$ vacua in the LVS. Hence, we have to consider at least two small cycles and a relating D-term constraint to construct $dS$ vacua.

We also have to consider a necessary condition for coexistence of the vanishing D-term constraint with the non-perturbative terms in superpotential. If the rigid divisors for the two non-perturbative effects intersect with the divisor on which the D-term constraint is generated via magnetic flux, we have to worry whether the VEV of matter fields that are generated by this magnetic flux are given accordingly such that the coefficients of non-perturbative terms remain non-zero. On the other hand, we may avoid additional zero mode contributions in a setup with minimal intersections. A general constraint is that the non-perturbative effects and D-term potential have to fulfill all known consistency condition, for instance requiring rigid divisors, avoiding Freed-Witten anomalies~\cite{Minasian:1997mm,Freed:1999vc} and saturating D3, D5, and D7 tadpole constraints. We expect these constraints to become less severe as the number of K\"ahler moduli increases beyond $h^{1,1}_+=3$ as in principle the degrees of freedom such as flux choices and rigid divisors increases.

This paper is organized as follows. We illustrate the uplift proposal generated through the multi-K\"ahler moduli dependence in the F-term potential and the required general geometric configuration in Section~\ref{genmech_sec} and give some computational details in Appendix~\ref{app_fluxconstraint}. We further discuss the applicability of the uplift mechanism in more general Swiss-Cheese type Calabi-Yau manifolds in Section~\ref{sec:real-more-moduli}.

\section{D-term generated racetrack uplift - general mechanism} \label{genmech_sec}

We illustrate the uplift mechanism by a D-term generated racetrack in Calabi-Yaus with the following properties: there are two small 4-cycles and two linear combinations of these small cycles that are rigid such that the existence of two non-perturbative terms is guaranteed in the superpotential avoiding additional fermionic zero modes from cycle deformations or Wilson lines. We show that this setup in general allows to stabilize the moduli in a $dS$ vacuum at large volume.

\subsection{Geometric setup and superpotential}

We consider an orientifolded Calabi-Yau $X_3$ with $h^{1,1}_+\geq3$ with the following general volume form of the divisors $D_i$
\begin{equation}
 \Vol = \frac16 \left(\sum_{i,j,k=1}^{h^{1,1}_+}\kappa_{i,j,k} t_i t_j t_k \right)\,,\label{gen2cycles}
\end{equation}
in terms of 2-cycle volumes $t_i$ and intersection numbers
\begin{equation}
\kappa_{ijk} =  \int_{X_3} D_i \wedge D_j \wedge D_k\,.
\end{equation}
The 4-cycle volumes are given as
 \begin{equation}
  \tau_i = \frac{\partial \Vol}{\partial t_i} = \frac12 \kappa_{ijk} t_j t_k\,.\label{tauandt}
 \end{equation}
We assume that $X_3$ has a  Swiss-Cheese structure with a big cycle named $D_a$ and at least two small cycles $D_b$ and $D_c$, i.e., its volume form can be brought to the form
\begin{equation}
 \Vol = \gamma_a \tau_a^{3/2} - \gamma_b \tau_b^{3/2}  - \gamma_c \tau_c^{3/2} - \Vol_{\text{rest}}\,, 
\end{equation}
with $\Vol_{\text{rest}}$ parametrizing the dependence of the volume on the remaining $h^{1,1}_+-3$ moduli. Now let us assume there are two rigid divisors $D_2$ and $D_3$ of which a linear combination forms the small cycles $D_b$ and $D_c$.
\begin{align}
\begin{aligned}
 D_2 &= d_{2b} D_b + d_{2c} D_c\,,\\
 D_3 &= d_{3b} D_b + d_{3c} D_c\,.
 \end{aligned}\label{tau2taua}
\end{align}
Even if there do not exist two divisors $D_2$ and $D_3$ that are rigid, one might still be able to effectively `rigidify' one or more divisors by fixing all the deformation moduli of the corresponding D7-brane stacks via a gauge flux choice~\cite{Bianchi:2011qh,Cicoli:2012vw,Louis:2012nb}. Under these assumptions, the superpotential in terms of the K\"ahler moduli $T_i = \tau_i + i\, \zeta_i$ is of the form
\begin{equation}
 W = W_0 + A_2 e^{-a_2 T_2} + A_3 e^{-a_3 T_3} = W_0 + A_2 e^{-a_2 \left(d_{2b} T_b + d_{2c} T_c \right)} + A_3 e^{-a_3 \left(d_{3b} T_b + d_{3c} T_c \right)}\,,\label{Wnonpert}
\end{equation}
with non-zero $A_2$, $A_3$ and $W_0$ being the Gukov-Vafa-Witten flux superpotential~\cite{Gukov:1999ya}.

\subsection{D7-brane and gauge flux configuration} \label{d7config3moduli_sec}

The orientifold plane O7 induces a negative D7 charge of $-8[O7]$ that has to be compensated by the positive charge of D7-branes. In general the O7 charge can be cancelled by introducing a Whitney brane with charge $8[O7]$~\cite{Collinucci:2008pf}. The non-perturbative effects of~\eqref{Wnonpert} can be either generated by ED3-instantons or gaugino condensation. For the latter, we choose a configuration with $N_2$ D7-branes on $D_2$ and $N_3$ D7-branes on $D_3$. In this case, the exponential coefficients of the non-perturbative terms in~\eqref{Wnonpert} are $a_2 = 2 \pi / N_2$ and $a_3 = 2 \pi / N_3$. The corresponding gauge group is either $SO(N)$ or $Sp(N)$ (which becomes $SU(N)$ if gauge flux is introduced), depending on if the divisor lies on the orientifold plane or not. Furthermore we introduce a third stack of $N_D$ branes on a general linear combination $D_D$ of basis divisors that is not either $D_2$ or $D_3$. This stack will introduce a D-term constraint that reduces the F-term effective scalar potential by one degree of freedom/ K\"ahler modulus. In the case of $h^{1,1}_+=3$ this corresponds to a two K\"ahler moduli LVS potential plus an uplift term that allows $dS$ vacua as we will show in Section~\ref{effectiveVgen_sec}.\footnote{In the case of $D_D$ being a linear combination of only $D_2$ and $D_3$ this divisor is only meaningful if $D_2$ and $D_3$ intersect as a linear combination of non-intersecting and rigid, i.e., local, four-cycles would not make sense. This is the reason we consider non-zero intersections between $D_2$ and $D_3$ in the first place as opposed to the more simple setup $\Vol \sim \tau_1^{3/2} - \tau_2^{3/2} - \tau_3^{3/2} - \Vol_{\text{rest}}$.} Note that in general all required D7-brane stacks have to be consistent with possible factorizations of the Whitney brane that cancels  the O7 charge~\cite{Collinucci:2008pf,Cicoli:2011qg}.

The D-term constraint is enforced via a Fayet-Illiopoulos (FI) term
\begin{equation}
 \xi_D = \frac{1}{\Vol} \int_{X_3} D_D \wedge J \wedge \mathcal{F}_D = \frac{1}{\Vol} q_{Dj} t_j\,,\label{FID}
\end{equation}
where $J=t_i D_i$ is the K\"ahler form on $X_3$ and $q_{Dj} = \tilde f^k_D \kappa_{Djk}$ is the anomalous $U(1)$-charge of the K\"ahler modulus $T_j$ induced by the magnetic flux $\mathcal{F}_D = \tilde f^k_D D_k$ on $D_D$. We choose flux-quanta $\tilde f^k_D$ such that $\xi_D = 0$ in~\eqref{FID} implies
\begin{equation}
 \tau_c = c\, \tau_b\,,\label{tauctaubprop}
\end{equation}
with a constant $c$ depending on flux quanta and triple intersection numbers. In a concrete example it is important to check that a constant $c$ in~\eqref{tauctaubprop} is realized which is consistent with stabilizing the moduli inside the K\"ahler cone of the manifold.

An important constraint arises from the requirement of two non-vanishing non-perturbative effects $A_2, A_3\neq 0$ on generally intersecting cycles $D_2$ and $D_3$. The cancellation of Freed-Witten anomalies requires the presence of fluxes $\mathcal{F}$ on the D7-branes wrapping these divisors that can potentially forbid the contribution from gaugino condensation in the superpotential. This gauge invariant magnetic flux $\mathcal{F}$ is determined by the gauge flux $F$ on the corresponding D7-brane and pull-back of the bulk $B$-field on the wrapped four-cycle via
\begin{equation}
 \mathcal{F} = F - B\,.
\end{equation}
If $D_2$ and $D_3$ intersect each other, the $B$-field can in general not be used to cancel both of theses fluxes to zero. However, it is still possible that both fluxes $\mathcal{F}_2$ and $\mathcal{F}_3$ can be chosen to be effectively trivial, such that no additional zero modes and FI-terms are introduced. These zero-modes would be generated via charged matter fields arising at the intersection of D7-brane stacks or from the bulk D7 spectrum. The constraint has to be checked on a case-by-case basis. We work out a sufficient condition on the intersections $\kappa_{ijk}$ for $\mathcal{F}_2$ and $\mathcal{F}_3$ to be trivial for the case of $D_2$ and $D_3$ not intersecting any other divisors $\kappa_{2,j,k} = \kappa_{3,j,k} =0$ for $j,k\neq 2,3$ in Appendix~\ref{app_fluxconstraint}. Furthermore, it has to be checked that $\mathcal{F}_D$ does not generate any additional zero-modes at the intersections of $D_D$ with $D_2$ and $D_3$.

Finally, the chosen D7-brane and gauge flux setup has to be consistent with D3, D5 and D7 tadpole cancellation. As for every explicit construction this has to be checked on a case-by-case basis for the particular manifold under consideration. We do not expect tadpole cancellation to be in general more restrictive than in e.g., the $AdS$ LVS. In particular, we do not require a large number of D7-branes~\cite{Westphal:2006tn} and/or racetrack effect~\cite{Sumitomo:2013vla} on a particular single divisor to achieve a large volume as in the K\"ahler Uplifting scenario.

\subsection{Effective potential of the K\"ahler moduli} \label{effectiveVgen_sec}

We start with a slightly simplified model where the F-term potential
\begin{equation}
 V_F = e^{K} \left( K^{\alpha\bar{\beta}} D_\alpha W \overline{D_\beta W} - 3 |W|^2 \right)\,,\label{VF}
\end{equation}
is given by
\begin{equation}
 \begin{split}
  K=& - 2 \ln \left({\cal V} + {\xi \over 2}\right), \quad
  {\cal V} = (T_a + \bar{T}_a)^{3/2} - (T_b + \bar{T_b})^{3/2} - (T_c + \bar{T_c})^{3/2},\\
  W =& W_0 + A_2 e^{-a_2 T_b} + A_3 e^{-a_3(T_b + T_c)},
 \end{split}
 \label{F-term potential-simplified}
\end{equation}
where we have used equal intersection numbers $\gamma$ and assumed stabilization of the dilaton and complex structure moduli via fluxes~\cite{Giddings:2001yu}. The values of these parameters are not essential for the uplift dynamics we illustrate in this paper. The superpotential in~\eqref{F-term potential-simplified} corresponds to a particular choice of the general linear combination in~\eqref{Wnonpert}. The model~\eqref{F-term potential-simplified} is known to include the solutions of the LVS~\cite{Balasubramanian:2005zx} that stabilizes the moduli in a non-supersymmetric way in the presence of the leading $\alpha'$-correction~\cite{Becker:2002nn} and one non-perturbative term. The $\alpha'$-correction is given by $\xi \propto - \chi g_s^{-3/2}$ where $\chi$ is the Euler number of the Calabi-Yau manifold.\footnote{Recently, it has been argued that the leading correction in both $\alpha'$ and string coupling constants on $SU(3)$ structure manifold comes with the Euler characteristic of the six-dimensional manifold as well as Calabi-Yau compactifications~\cite{Grana:2014vva}.}

The D-term potential is given through the magnetized D7-branes wrapping the Calabi-Yau divisor $D_i$~\cite{Haack:2006cy}:
\begin{equation}
 V_D = {1 \over \re (f_D)} \left(\sum_j c_{Dj} \hat{K}_j \varphi_j - \xi_D\right)^2\,,
  \label{D-term potential}
\end{equation}
where the gauge kinetic function
\begin{equation}
 \text{Re}\,(f_D) = \frac12\, \int_{D_D} J \wedge J - \frac{1}{2g_s}\int_{D_D} \mathcal{F}_D \wedge \mathcal{F}_D \,,\label{fD}
\end{equation}
and $\varphi_j$ are matter fields associated with the diagonal $U(1)$ charges $c_{Dj}$ of a stack of D7-branes and the FI-term $\xi_D$ is defined in~\eqref{FID}.

Now we redefine the coordinates:
\begin{equation}
 T_s \equiv {1\over 2 }\left(T_b + T_c  \right), \quad Z \equiv {1\over 2} \left(T_b - T_c \right).
\end{equation}
When the D7-branes wrapping the divisor $D_D$ are magnetized and the matter fields are stabilized either at $\langle \varphi_i \rangle = 0$ or satisfying $ \langle \sum c_{ij} \hat{K}_j \varphi_j \rangle =0$, the D-term potential may become
\begin{equation}
 V_D \propto {1\over \re(f_D)}{1 \over {\cal V}^2} \left(\sqrt{\tau_b} - \sqrt{\tau_c} \right)^2\,,
  \label{D-term potential in simple model}
\end{equation}
using $\xi_D \propto \sqrt{\tau_b} - \sqrt{\tau_c}$ implied by the flux $\mathcal{F}_D$, see~\eqref{tauctaubprop} where we use $c=1$ for simplicity. In the large volume limit, the F-term potential generically scales as ${\cal O} ({\cal V}^{-3})$ in the minima given in the LVS model. Stabilizing the K\"ahler moduli at ${\cal O} ({\cal V}^{-3})$ then requires a vanishing D-term potential, i.e., $\tau_b = \tau_c$ corresponding to $z \equiv \re Z=0$.

Thanks to the topological coupling to the two-cycle supporting magnetic flux, the imaginary mode of the $Z$ modulus is eaten by a massive $U(1)$ gauge boson through the St\"uckelberg mechanism. Since the gauge boson has a mass of order of the string scale ${\cal O} ({\cal V}^{-1/2})$, the degree of freedom of $\im Z$ charged under the anomalous $U(1)$ as well as the gauge boson is integrated out at the high scale. Hence, we are left with the stabilization of the remaining moduli fields by the F-term potential.

\subsection{F-term uplift\label{sec:f-term-uplift}}

Next we will consider the stabilization by the F-term potential given in (\ref{F-term potential-simplified}). We are interested in LVS like minima ${\cal V} \sim e^{\hat{a}_i \tau_i}$ realizing an exponentially large volume. Then the leading potential of order ${\cal V}^{-3}$ is given by
\begin{equation}
 \begin{split}
  V & \sim {3 W_0^2 \xi \over 4 {\cal V}^3} + {2 W_0 \over {\cal V}^2} \left( a_2 A_2 {\tau_b} e^{-a_2 \tau_b/2} + a_3 A_3 (\tau_b + \tau_c) e^{-a_3 (\tau_b + \tau_c)/2}\right) \\
  & + {2 \over 3 {\cal V}} \left(a_2^2 A_2^2 \sqrt{\tau_b} e^{- a_2 \tau_b} + a_3^2 A_3^2 (\sqrt{\tau_b}+\sqrt{\tau_c}) e^{-a_3 (\tau_b + \tau_c)} + 2 a_2 a_3 A_2 A_3 \sqrt{\tau_b} e^{-a_2 \tau_b/2 - a_3 (\tau_b + \tau_c)/2} \right)\,,
 \end{split}
 \label{effective potential at larger volume 0}
\end{equation}
where the imaginary directions are stabilized at $\im T_i= 0$, and $\im T_a$ is stabilized by non-perturbative effects that are omitted in (\ref{F-term potential-simplified}), inducing a very small mass for $\im T_a$ and with negligible influence on the stabilization of the other moduli. Although the general minima of $\im T_i$ are given by $a_i \im T_i = m_i \pi$ with $m_i \in {\mathbb Z}$, the different solutions just change the sign of $A_i$ and thus we can simply have the potential of the above form.

As the D-term stabilizes $\tau_c = \tau_b$, the resultant potential becomes 
\begin{equation}
 \begin{split}
  V &\sim {3 W_0^2 \xi \over 4 {\cal V}^3} + {4 W_0 \over {\cal V}^2} \left(a_2 A_2 \tau_s e^{-a_2 \tau_s} + 2 a_3 A_3 \tau_s e^{-2 a_3 \tau_s}\right)\\
    &+ {2 \sqrt{2} \over 3 {\cal V}} \left({a_2^2 A_2^2 \sqrt{\tau_s}} e^{-2 a_2 \tau_s} + {4 a_3^2 A_3^2 \sqrt{\tau_s}} e^{-4 a_3 \tau_s} + 2 a_2 a_3 A_2 A_3 \sqrt{\tau_s} e^{-(a_2 + 2 a_3)\tau_s} \right)\,,
 \end{split}
 \label{effective potential at larger volume with ts}
\end{equation}
where we have defined $\tau_s = \re T_s$. One may consider that this form of the potential looks similar to the racetrack type. Although cross terms of $A_2, A_3$ appear due to the $T_b$ dependence, the important point for the uplift mechanism demonstrated in this paper is that the cross terms between $T_b$ dependence of the $A_2$ term and $T_c$ dependence of the $A_3$ term appear at ${\cal O} ({\cal V}^{-4})$.\footnote{Note that this would be more obvious if we start from a toy setup with $W = W_0 + A_2 e^{-a_2 T_b} + A_3 e^{-a_3 T_c}$, although one might not obtain the D-term constraint like $\tau_c = c \tau_b$.} If the cross term appears at ${\cal O}({\cal V}^{-3})$, it disturbs uplifting to $dS$.

We further redefine the fields and parameters such that there are no redundant parameters affecting the stabilization:
\begin{equation}
 \begin{split}
  &x_s = a_2 \tau_s, \quad {\cal V}_x = {\cal V} a_2^{3/2},\\
  &c_i = {A_i \over W_0},\quad  \xi_x = \xi a_2^{3/2}, \quad \beta = {2 a_3 \over a_2}.
 \end{split}
 \label{redefined parameters}
\end{equation}
Then the effective potential at order ${\cal O}({\cal V}^{-3})$ becomes
\begin{equation}
 \begin{split}
  \hat{V} \equiv \left({a_2^{-3} W_0^{-2}}\right) V \sim {3\xi_x \over 4 {\cal V}_x^3} + {4 c_2 x_s \over {\cal V}_x^2}e^{-x_s} + {2\sqrt{2} c_2^2 x_s^{1/2}\over 3 {\cal V}_x} e^{-2 x_s } + {4 \beta c_3 x_s \over {\cal V}_x^2} e^{-\beta x_s}.
 \end{split}
 \label{redefined potential}
\end{equation}
We have neglected the term proportional to $c_3^2$ and the cross term between $c_2, c_3$ in the expression above. In fact, these terms are not important for the uplift mechanism of our interest, and we will justify this assumption a posteriori later. Since the uplift term comes together with $e^{-\beta x_s}$, this term contributes of ${\cal O}({\cal V}_x^{-3})$ when $\beta \sim {\cal O} (1)$. Hence, it contributes at the same order as the stabilizing F-term potential and no suppression factor provided by warping or dilaton dependence is required.

Before performing the uplift, we consider the LVS solution by setting $c_3 =0$. We use a set of parameters:
\begin{equation}
 c_2 = -0.01, \quad \xi_x = 5.
  \label{test input parameter}
\end{equation}
The extremal equations $\partial_I \hat{V} = 0$  at $c_3=0$ can be simplified as
\begin{equation}
 \xi_x = {64 \sqrt{2} (x_s-1) x_s^{5/2} \over (4 x_s -1)^2}, \quad
  c_2 = - {6\sqrt{2} (x_s-1) x_s^{1/2} \over (4 x_s -1) {\cal V}_x} e^{-x_s}.
\end{equation}
Solving the equations above, we obtain
\begin{equation}
 {\cal V}_x \sim 467, \quad x_s \sim 1.50.\label{VxxsAds}
\end{equation}
We can easily check that this solution gives an $AdS$ vacuum. Note that when we have just two moduli fields ${\cal V}_x, x_s$ in the LVS, the positivity of $\xi_x$ automatically guarantees the stability of the minima since the required condition $x_i > 1$ is satisfied (see e.g.~\cite{Rummel:2013yta}).

\begin{figure}[t]
\includegraphics[width=20.5em]{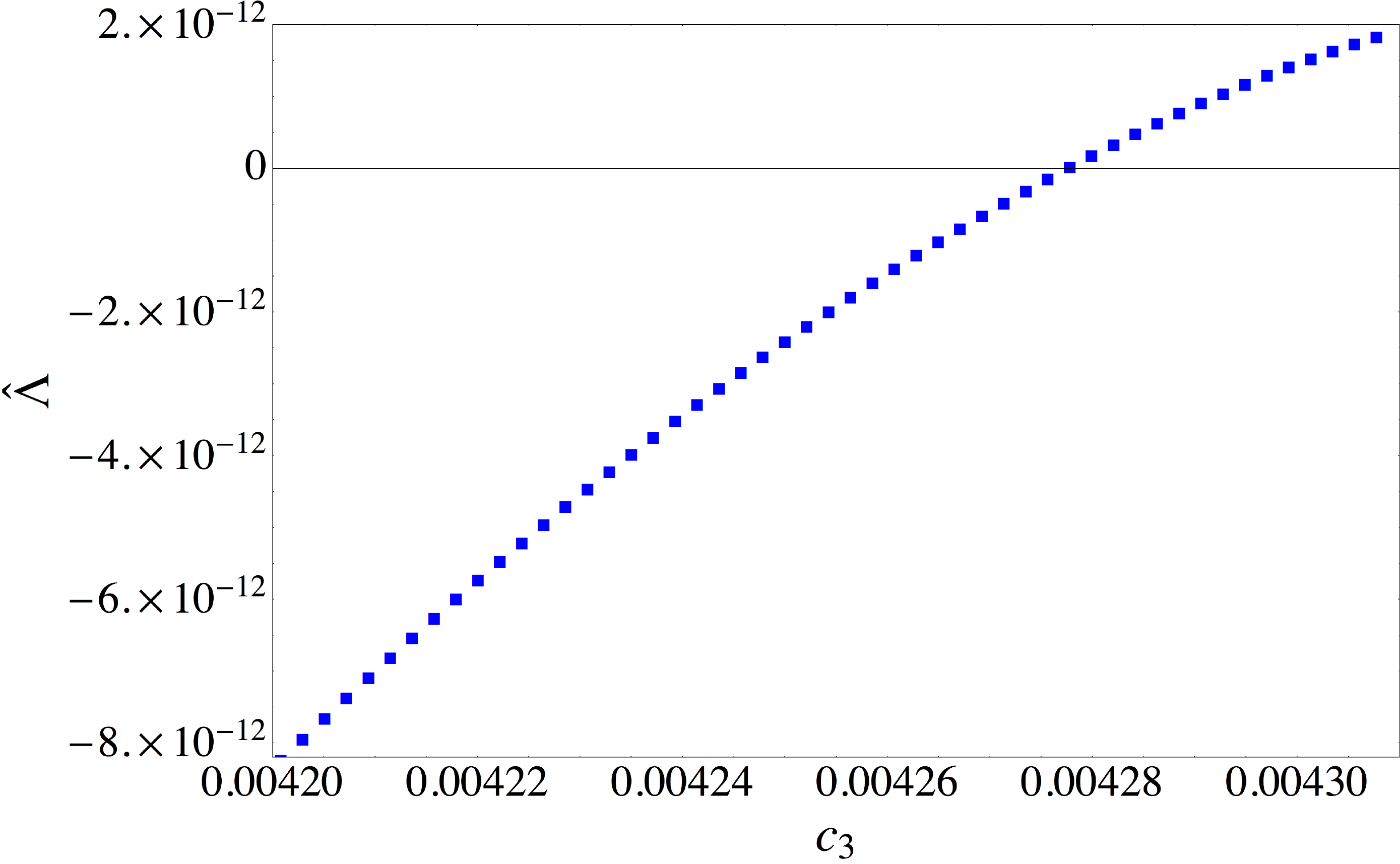}
\vspace{4mm}
\includegraphics[width=20em]{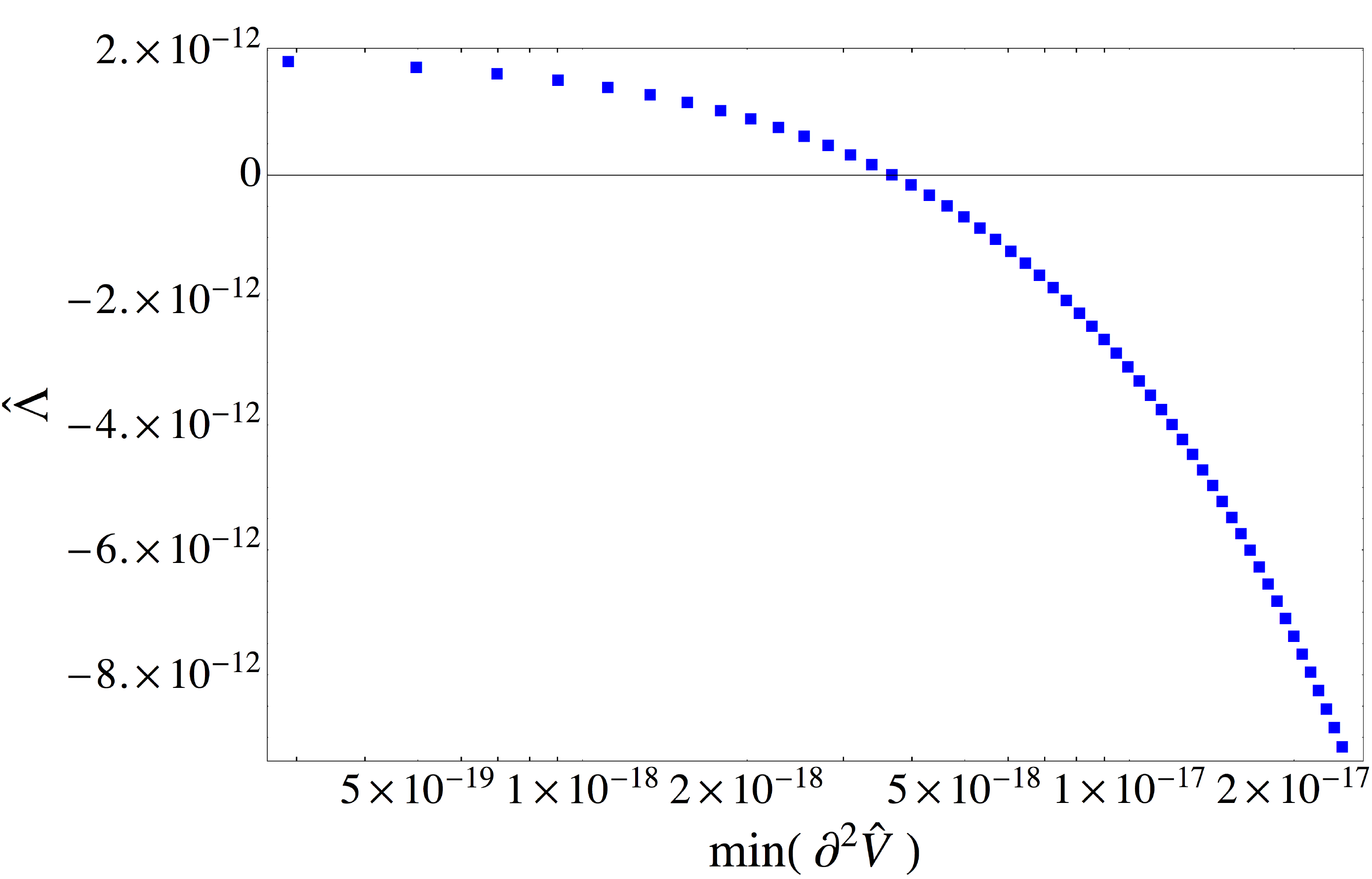}
\includegraphics[width=20.5em]{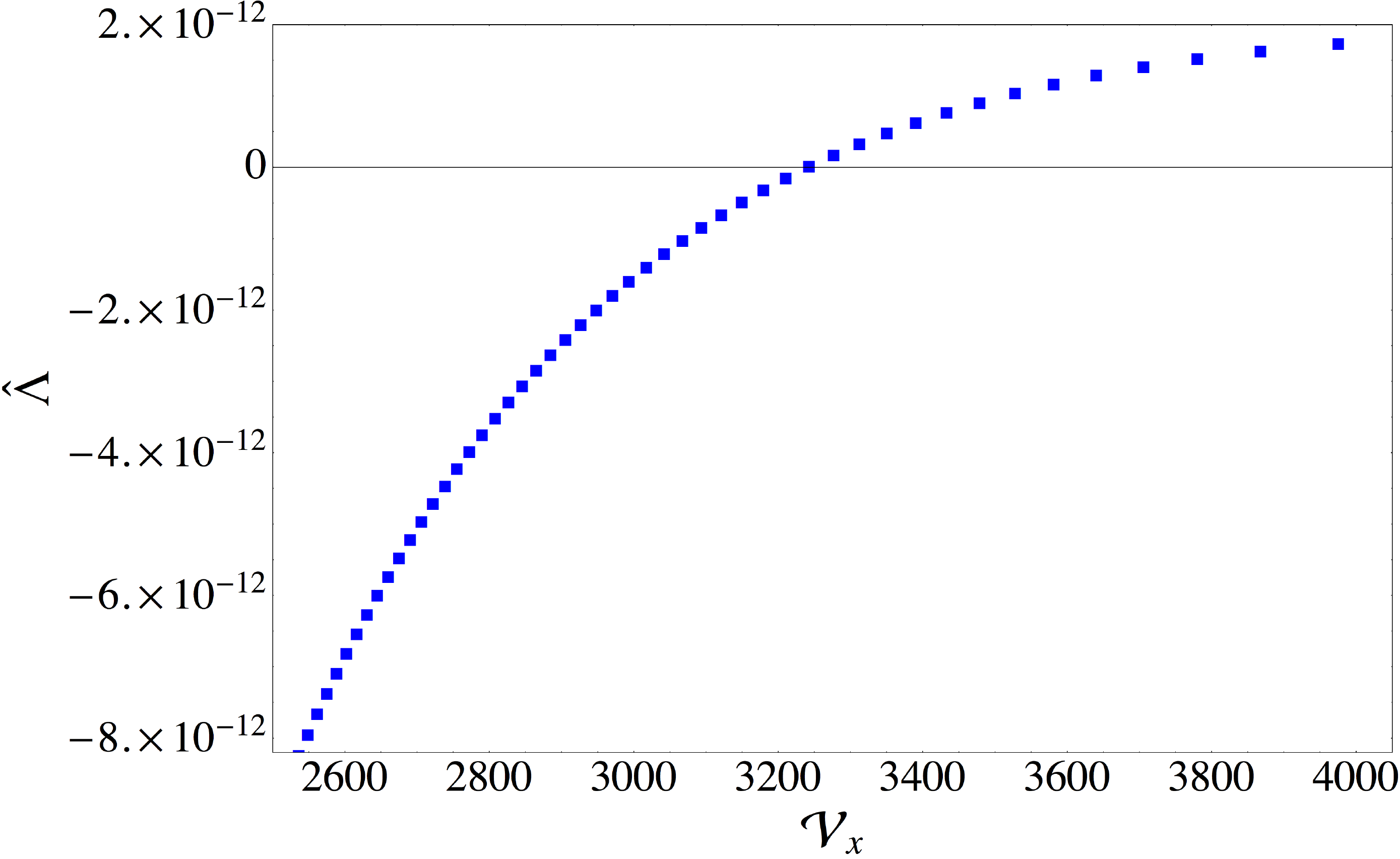}
\caption{\footnotesize Illustration of the D-term generated racetrack uplift mechanism. We plot the cosmological constant $\hat \Lambda$ vs $c_3$, $\text{min}(\partial^2 \hat V)$ and $\Vol_x$ at the minima of the potential, especially near the Minkowski point.}
\label{fig:uplift-illustration}
\end{figure}

Now we consider non-zero $c_3$ for the uplift. As $c_3$ increases, the vacuum energy of the potential minimum increases and eventually crosses the Minkowski point. In Figure~\ref{fig:uplift-illustration}, we illustrate the behavior of the minimum point by changing the value of $c_3$. Interestingly, the volume increases as the vacuum energy increases, suggesting that the effective description of the theory will be more justified toward $dS$ vacua. On the other hand, the minimum value of the Hessian decreases. Destabilization occurs when the uplift term dominates the entire potential. As this happens at higher positive values of the cosmological constant, there certainly exist a range of parameters yielding stable $dS$ vacua within this setup.

As a reference, we show numerical values of parameters close to crossing the Minkowski point. When we use
\begin{equation}
 \beta = {5\over 6},
  \label{test beta value}
\end{equation}
the minimum reaches Minkowski at
\begin{equation}
 c_3 \sim 4.28 \times 10^{-3}, \quad {\cal V}_x \sim 3240, \quad x_s \sim 3.07.
\end{equation}
So we see that the volume increases quite drastically from the $AdS$ vacuum~\eqref{VxxsAds}. Since $c_3$ remains small compared to the input value of $c_2$, we see that our approximation neglecting the term proportional to $c_3^2$ is justified.

In fact, it is not difficult to see how these values change when the presence of $c_3^2$ terms and cross terms between $c_2,c_3$ in the potential (\ref{effective potential at larger volume with ts}) are taken into account. With the input parameters we used in (\ref{test input parameter}), the Minkowski vacuum is obtained when
\begin{equation}
 c_3 \sim 5.11 \times 10^{-3}, \quad {\cal V}_x \sim 2860, \quad x_s \sim 2.61.
  \label{test solution with c_3^2}
\end{equation}
Since the obtained values are not significantly different from the case where $c_3^2$ terms and cross terms between $c_2, c_3$ are neglected, we conclude a posteriori that the uplift term is dominated by the term linear in $c_3$.

Let us comment on the stabilization of the axionic partner of each modulus field. As stated, the imaginary mode of $Z$ is eaten up by a massive gauge boson and hence integrated out at the high scale. The axionic partner of the big divisor $T_a$ is stabilized by non-perturbative effects yielding a tiny mass. The remaining modulus $\im T_s$ is stabilized by the F-term potential as the D-term potential does not depend on the latter. In the approximated potential up to ${\cal O}(\Vol^{-3})$, the Hessian of $ y_s = a_2 \im T_s$ is
\begin{equation}
\partial_{y_s}^2 \hat{V}|_{\rm ext} \sim  5.14 \times 10^{-10}\,,
\end{equation}
where we have included $c_3^2$ and $c_2, c_3$ cross terms, and used the solution (\ref{test solution with c_3^2}) and $\im T_{a} = \im T_i =0$. Thus all K\"ahler moduli are stabilized.

\subsection{Analytical estimate}

It is difficult to analytically derive a generic condition for the D-term generated racetrack uplift since the formulas are still complicated enough even after using several approximations. However, some of the expressions can be simplified under an additional reasonable approximation. In this subsection, we illustrate some analytical analyses for a better understanding.

Since we checked that the uplift mechanism works even at linear approximation of the uplift parameter $c_3$, we only keep terms up to linear order in $c_3$ and neglect the higher order terms including cross terms. The extremal condition $\partial_i \hat{V} = 0$ of the potential (\ref{redefined potential}) is now simplified by
\begin{equation}
 \begin{split}
  &c_2 \sim -{6\sqrt{2x_s} (x_s-1) \over 4 x_s -1} {e^{x_s} \over {\cal V}_x} + c_3 {\beta (\beta x_s -1) \over x_s -1} e^{(1-\beta)x_s} ,\\
  &\xi_x \sim {64\sqrt{2} x_s^{5/2} (x_s-1) \over (4x_s-1)^2} - c_3 {32 \beta x_s^2 \left(2(\beta+2)x_s + \beta-7 \right) \over 9 (x_s-1) (4 x_s-1)} e^{-\beta x_s} {\cal V}_x .
 \end{split}
 \label{extremal condition of effective potential}
\end{equation}
Although our interest is the uplift toward $dS$ vacua, we have to cross the Minkowski point along the way. Thus, the condition that the minimum structure holds when uplifted to Minkowski vacua is a necessary condition for the $dS$ uplift mechanism. The condition for Minkowski at the extrema $\hat{V}|_{\rm ext} = 0$ reads
\begin{equation}
 \begin{split}
  c_3 \sim {18 \sqrt{2} (\beta-1) x_s^{3/2} \over \beta (4 (\beta-1) x_s - 3)^2} {e^{\beta x_s} \over {\cal V}_x}.
 \end{split}
 \label{minkowski condition}
\end{equation}

Next, we proceed to check the stability at the Minkowski point. Although we know the conditions to check the stability, the formula of the Hessian is yet too complicated to perform an analytical analysis. So we further focus on the region satisfying $x_s \gg 1$. The region with $x_s \gg 1$ is motivated since the $AdS$ minimum points, before adding an uplift term, are guaranteed to have a positive Hessian since all eigenvalues are positive definite when satisfying $x_s > 1$ in LVS type stabilizations (see e.g.~\cite{Rummel:2013yta}). Furthermore, higher instanton corrections can be safely neglected. As shown in Figure~\ref{fig:uplift-illustration}, the minima can be uplifted keeping the positivity of the Hessian until reaching the destabilization point with a relatively high positive vacuum energy. Hence, having $x_s \gg 1$ is motivated to see the basic feature of the D-term generated racetrack uplift mechanism. Since there is no reason to take $\beta$ to be small/large, we consider $\beta \sim {\cal O} (1)$.

When we use the approximation $x_s \gg 1$, a component of the Hessian and the determinant at the extrema become (\ref{extremal condition of effective potential})
\begin{equation}
 \begin{split}
  \partial_{{\cal V}_x}^2 \hat{V}|_{\rm ext} \sim& {6\sqrt{2} x_s^{3/2} \over {\cal V}_x^5} - c_3 {8 \beta x_s e^{-\beta x_s} \over {\cal V}_x^4}
  \sim {6\sqrt{2} x_s^{3/2} \over {\cal V}_x^5},\\
  \det \left(\partial_i \partial_j \hat{V} \right)|_{\rm ext} \sim& {162 x_s^2 \over {\cal V}_x^8} + c_3 {24\sqrt{2} \beta(\beta^2 + \beta -2) x_s^{5/2} e^{-\beta x_s} \over {\cal V}_x^7}
  \sim {54 (1- \beta) x_s^2 \over {\cal V}_x^8}\,,
 \end{split}
 \label{analytical hessian at minkowski}
\end{equation}
where in the last step of both equations, we have used the Minkowski condition (\ref{minkowski condition}). According to Sylvester's criterion, the positivity of a matrix can be checked by the positivity of the determinant of all sub-matrices. Thus it is enough to check the positivity of the quantities in (\ref{analytical hessian at minkowski}). Therefore we conclude that the stability at the Minkowski point requires $\beta < 1$. This condition is clearly satisfied in the previous numerical example following (\ref{test beta value}), which may justify the crude approximations we took in this subsection. Note that the Hessian of the imaginary mode is guaranteed to be positive under the above used approximations:
\begin{equation}
 \partial_{y_s}^2 \hat{V}|_{\rm ext} \sim {6\sqrt{2} x_s^{3/2} \over {\cal V}_x^3} - c_3 {4 \beta^2 (\beta +1) x_s e^{-\beta x_s} \over {\cal V}_x^2}
  \sim {6\sqrt{2} x_s^{3/2} \over {\cal V}_x^3}.
\end{equation}

Finally, let us check the extremal and Minkowski conditions in the limit $x_s \gg 1$. Now all conditions are simplified to be
\begin{equation}
 \begin{split}
  \xi_x \sim 4 \sqrt{2} x_s^{3/2}, \quad
  c_2 \sim -{3 \sqrt{x_s} \over \sqrt{2}} {e^{x_s} \over {\cal V}_x}, \quad
  c_3 \sim {9 \over 4 \sqrt{2x_s} \beta (1-\beta)} {e^{\beta x_s} \over {\cal V}_x}.
 \end{split}
\end{equation}
We see that the minimum point needs $\xi_x >0$ and $c_2 < 0$ in agreement with the minimum requirement of the two-moduli LVS at $AdS$. The stability condition $\beta < 1$ suggests $c_3 > 0$. In fact, the extremal condition for $\xi_x, c_2$ is simply the leading order approximation of each first term in \eqref{extremal condition of effective potential} as the $c_3$ contribution appears sub-dominant. This justifies that the linear approximation for $c_3$ is compatible with $x_s \gg 1$. Hence, we can regard the last term in the potential (\ref{redefined potential}) as the uplift term.

\section{On realization in models with more moduli\label{sec:real-more-moduli}}

In this section, we show that the uplift mechanism works well in the presence of additional K\"ahler moduli in Swiss-Cheese type Calabi-Yau compactifications. We consider a simple toy model with $h^{1,1}_+=4$, which captures the essential features of the D-term generated racetrack uplift mechanism defined by
\begin{equation}
 \begin{split}
  K=& - 2 \ln \left({\cal V} + {\xi \over 2}\right), \quad
  {\cal V} = (T_a + \bar{T}_a)^{3/2} - (T_b + \bar{T_b})^{3/2} - (T_c + \bar{T_c})^{3/2} - (T_e + \bar{T_e})^{3/2},\\
  W =& W_0 + A_2 e^{-a_2 T_b} + A_3 e^{-a_3 (T_b+T_c)} + A_4 e^{- a_4 T_e}.
 \end{split}
\end{equation}
Again we are interested in the case of a Swiss-Cheese volume for moduli stabilization of the LVS type. Note that we used the name $T_e$ to avoid confusion with $T_D$.

Taking into account the D-term potential generated by the magnetized D7-branes wrapping the divisor $Z$, we assume again that the $Z =\frac12 (T_b - T_c)$ modulus is stabilized at $Z=0$. Setting $a_4 = a_2$ for simplicity, the effective potential at ${\cal V}^{-3}$ from the F-terms is given by
\begin{equation}
 \begin{split}
   {\hat{V}} \equiv \left({a_2^{-3} W_0^{-2}}\right) V \sim& {3\xi_x \over 4 {\cal V}_x^3} + {4 c_2 x_s \over {\cal V}_x^2}e^{-x_s} + {2\sqrt{2} c_2^2 x_s^{1/2}\over 3 {\cal V}_x} e^{-2 x_s}
  + {4 c_4 x_4 \over {\cal V}_x^2}e^{-x_4} + {2\sqrt{2} c_4^2 x_4^{1/2}\over 3 {\cal V}_x} e^{-2 x_4}\\
  &+ {4 \beta c_3 x_s \over {\cal V}_x^2} e^{-\beta x_s} + {\sqrt{2} c_3^2 x_s^{1/2}\over 3 {\cal V}_x} e^{-2 \beta x_s} + {2\sqrt{2}\beta c_2 c_3 x_s^{1/2} \over 3 {\cal V}_x} e^{-(1+\beta)x_s},
 \end{split}
\end{equation}
where we have further defined $x_e = a_4 \tau_e, \beta=2a_3/a_2$ and $c_4 = A_4/ W_0$ in addition to~\eqref{redefined parameters}. Here we included the term proportional to $c_3^2$ as well as the cross term $c_2 c_3$ even though they are potentially subleading.

When we use a set of parameters:
\begin{equation}
 c_2 = -0.0114, \quad c_4 = -3.38 \times 10^{-4}, \quad \xi_x = 19,
  \label{input of 4 moduli model}
\end{equation}
then the $AdS$ LVS minimum at $c_3=0$ is located at
\begin{equation}
 \begin{split}
  {\cal V}_x \sim 2740, \quad x_s \sim 2.60, \quad x_e \sim  1.12.
 \end{split} \label{4modminval}
\end{equation}
The stability of multi-K\"ahler moduli models of the LVS type is ensured if the constraint $x_i>1$ is satisfied~\cite{Rummel:2013yta}. Hence, the extremal point~\eqref{4modminval} is stable.

Now we add the uplift terms $c_3\neq 0$ and $\beta=5/6$. The minimum with the input parameters (\ref{input of 4 moduli model}) reaches Minkowski at
\begin{equation}
 \begin{split}
  c_3 \sim 4.55 \times 10^{-3}, \quad {\cal V}_x \sim 5.64 \times 10^4, \quad x_s \sim 5.45, \quad x_e \sim 2.26.
 \end{split}
\end{equation}
Although the volume is drastically changed during the uplift toward $dS$ vacua, we can check the stability of the minimum by plugging the values into the Hessian, similarly to the simple three moduli model. The cosmological constant can further increase in the positive region keeping the stability until the minima exceeds the potential barrier where decompactification happens.

\section{Discussion}

We have proposed an uplift mechanism using the structure of at least two small K\"ahler moduli $T_b$ and $T_c$ in Swiss-Cheese type compactifications. The uplift contribution arises as an F-term potential when using a D-term condition which fixes $\re T_b = c \re T_c$ at a higher scale, where $c$ is determined by magnetized fluxes on D7-branes. The uplift term becomes of the form $e^{-a_s \tau_s}/{\cal V}^2$ at large volumes, and hence it can naturally balance with the stabilizing potential in the Large Volume Scenario (LVS), without requiring suppressions in the coefficient, for instance, by warping or a dilaton dependent non-perturbative effect.

In addition, we have shown that the D-term generated racetrack uplift works in the presence of additional K\"ahler moduli. Together with the fact that constraints on the uplift parameters are rather relaxed, i.e., $\beta< 1$ and $c_3>0$, this makes us optimistic that there should be many manifolds admitting the proposed uplift mechanism.

Since the proposed uplift mechanism requires certain conditions for a D-term constraint and two non-vanishing non-perturbative effects, it should be interesting if we can construct an explicit realization of this model in a particular compactification. Such an explicit construction requires to match all known consistency conditions such as cancellation of Freed-Witten anomalies and cancellation of the D3, D5, and D7 tadpole~\cite{Cicoli:2011qg,Cicoli:2012vw,Louis:2012nb,Cicoli:2013mpa,Cicoli:2013zha}. We hope to report on an explicit example in another paper.

Furthermore, the phenomenological aspect of the proposed uplift mechanism should be interesting. Even though the moduli are essentially stabilized as in the LVS, the resultant behavior of the mass spectrum and/or soft SUSY breaking terms may be different depending on which uplift mechanism we employ to realize the $dS$ vacuum.

Finally, in this paper, we concentrated on analyzing the structure of $dS$ minima. However, the structure of the potential is also changed by the uplift term in regions that might be important for including inflationary dynamics. We relegate the analysis of possible inflation scenarios as well as phenomenological consequences compared to other uplift proposals to future work.

\section*{Acknowledgments}
We would like to thank Joseph P. Conlon and Roberto Valandro for valuable discussions and important comments, and also the organizers of the workshop "String Phenomenology 2014" held at ICTP Trieste, Italy, where some of the results of this paper were presented. YS is grateful to the Rudolph Peierls Centre for Theoretical Physics, University of Oxford where part of this work was done for their hospitality and support. MR is supported by the ERC grant `Supersymmetry Breaking in String Theory'. This work is partially supported by the Grant-in-Aid for Scientific Research (No. 23244057) from the Japan Society for the Promotion of Science.

\appendix
\section{Conditions for avoiding additional zero-modes} \label{app_fluxconstraint}

In this appendix, we give a sufficient condition on the intersections $\kappa_{ijk}$ for $\mathcal{F}_2$ and $\mathcal{F}_3$ to be trivial for the case of $D_2$ and $D_3$ not intersecting any other divisors $\kappa_{2,j,k} = \kappa_{3,j,k} =0$ for $j,k \neq 2,3$. This is a necessary condition for the non-perturbative effects on $D_2$ and $D_3$ to contribute to the superpotential, which is crucial for the uplift mechanism considered in this work.

In order to avoid Freed-Witten anomalies the gauge flux on the D7-branes has to satisfy
\begin{equation}
 F + \frac{c_1(D)}{2} \in H^2(X_2,\mathbb{Z})\,.
\end{equation}
In particular, if $D$ is non-spin, i.e., $c_1(D)$ is odd, $F$ is always non-zero. Using $c_1(D)=-D$, the magnetic fluxes on $D_2$ and $D_3$ become
\begin{equation}
 F_i = f_i^2 D_2 + f_i^3 D_3 + \frac{D_i}{2} \quad \text{with} \quad f_i^k \in \mathbb{Z} \quad \text{for} \quad i,k=2,3\,.
\end{equation}
Since $D_2$, $D_3$ and $D_D$ are all intersecting we have only one choice for the $B$-field to cancel one $\mathcal{F}$. We choose the $B$-field without loss of generality such that $\mathcal{F}_2 = F_2 - B = 0$. In this case, we get
\begin{equation}
 \mathcal{F}_3 = F_3 - F_2 = \left(f_3^2 - f_2^2 -\frac12\right)  D_2 + \left(f_3^3 - f_2^3 +\frac12 \right) D_3\,.
\end{equation}

In order to avoid additional FI-terms and/or zero-modes via chiral matter at brane intersections or in the bulk spectrum of the D7-brane stacks on $D_2$ and $D_3$, we have to demand the magnetic fluxes $\mathcal{F}_2$ and $\mathcal{F}_3$ to be effectively trivial which is the case for
\begin{equation}
 0 = \int_{D_i} \mathcal{F}_i \wedge J = \int_{X_3} J \wedge D_i \wedge \mathcal{F}_i \qquad \text{for} \quad i=2,3\,,\label{FIvanish23}
\end{equation}
for the K\"ahler form $J = t_1 D_1 + t_2 D_2 + t_3 D_3$. This condition is trivially fulfilled for the zero flux $\mathcal{F}_2$. For $\mathcal{F}_3$, \eqref{FIvanish23} becomes
\begin{align}
 \begin{aligned}
  0 &= \int_{X_3} J \wedge D_3 \wedge \mathcal{F}_3\,,\\ 
   &= \int_{X_3} (t_1 D_1 + t_2 D_2 + t_3 D_3) \wedge D_3 \wedge \left[  \left(f_3^2 - f_2^2 -\frac12\right)  D_2 +  \left(f_3^2 - f_2^2 +\frac12\right) D_3\right]\,,\\
   &= t_2 \left[ \left(f_3^2 - f_2^2 -\frac12\right)\kappa_{223} +  \left(f_3^2 - f_2^2 +\frac12\right)\kappa_{233} \right] + t_3 \left[ \left(f_3^2 - f_2^2 -\frac12\right)\kappa_{233} +  \left(f_3^2 - f_2^2 +\frac12\right)\kappa_{333} \right]\,,\label{F3det}
 \end{aligned}
\end{align}
using the intersection form~\eqref{gen2cycles}. For general $t_2,t_3\neq 0$ \eqref{F3det} is fulfilled if
\begin{equation}
 \frac{2 f^2_3 - 2 f_2^2 - 1}{2 f^3_3 - 2 f_2^3 + 1} = -\frac{\kappa_{333}}{\kappa_{233}} = -\frac{\kappa_{233}}{\kappa_{223}}\,,\label{fluxcond}
\end{equation}
where the last condition can be rewritten as $\kappa_{233}^2 = \kappa_{223} \kappa_{333}$. Clearly, \eqref{fluxcond} can not be fulfilled for general intersection numbers. The intersections that can accommodate the condition of trivial gauge flux $\mathcal{F}_2$ and $\mathcal{F}_3$ \eqref{fluxcond} are the following:
\begin{center}
\begin{itemize}
 \item $\kappa_{223}=\kappa_{233}=\kappa_{333}\neq0$ or
 \item $\kappa_{222}=\kappa_{223}=\kappa_{233}\neq0$ or
 \item $\kappa_{233} = Z^n$, $\kappa_{223} = Z^m$, $\kappa_{333} = Z^k$ with $Z$ being an odd integer and integers $k+m=2n$ or
 \item $\kappa_{223} = Z^n$, $\kappa_{222} = Z^m$, $\kappa_{233} = Z^k$ with $Z$ being an odd integer and integers $k+m=2n$,
\end{itemize} 
\end{center}
i.e., for either of these conditions there exist flux quanta $f_i^k$ such that $\mathcal{F}_2$ and $\mathcal{F}_3$ are trivial. The second and fourth condition stem from choosing the $B$-field such that $\mathcal{F}_3=0$.

\bibliographystyle{utphys}
\bibliography{myrefs}

\end{document}